\documentstyle[11pt,epsfig]{article}
\newcommand \be{\begin{equation}}
\newcommand \bea{\begin{eqnarray} \nonumber }
\newcommand \ee{\end{equation}}
\newcommand \eea{\end{eqnarray}}

\begin{document}

\title{{\Huge Stock Market Speculation}: \\Spontaneous Symmetry Breaking of Economic Valuation}

\author{Didier Sornette$^{1,2,3}$\\
$^1$ Institute of Geophysics and
Planetary Physics\\ University of California, Los Angeles, California 90095\\
$^2$ Department of Earth and Space Science\\
University of California, Los Angeles, California 90095\\
$^3$ Laboratoire de Physique de la Mati\`{e}re Condens\'{e}e\\ CNRS UMR6622 and
Universit\'{e} de Nice-Sophia Antipolis\\ B.P. 71, Parc
Valrose, 06108 Nice Cedex 2, France \\
e-mails: sornette@moho.ess.ucla.edu}

\date{\today}
\maketitle

\begin{abstract}

Firm foundation theory estimates a security's firm fundamental value
based on four determinants: expected growth rate, expected dividend payout,
the market interest rate and the degree of risk. In contrast, other views of decision-making
in the stock market, using alternatives such
as human psychology and behavior, bounded rationality, 
agent-based modeling and evolutionary game theory, 
expound that speculative and crowd behavior
of investors may play a major role in shaping market prices. Here, we propose that 
the two views refer to two classes of companies 
connected through a ``phase transition''. Our theory is based on 1)
the identification of the fundamental parity symmetry of prices ($p \to -p$), which
results from the relative direction of payment flux compared to
commodity flux and 2) the observation 
that a company's risk-adjusted growth rate discounted by the market
interest rate behaves as a control parameter for the 
observable price. We find a critical value of this control parameter at which
a spontaneous symmetry-breaking of prices occurs, leading to a spontaneous
valuation in absence of earnings, similarly to the emergence of a spontaneous
magnetization in Ising models in absence of a magnetic field. The low growth
rate phase is described by the firm foundation theory while the large growth
rate phase is the regime of speculation and crowd behavior. In practice, while
large ``finite-time horizon'' effects round off the predicted singularities,
our symmetry-breaking speculation theory accounts for the 
apparent over-pricing and the high volatility of fast growing companies on the stock markets.

\end{abstract}
\vskip 1cm

\section{Introduction}

There is a growing interest in understanding
the divergence in the stock market between``New Economy'' and
``Old Economy'' stocks, between technology and almost everything else. 
Over the past two
years, stocks in the Standard \& Poor's technology sector have risen nearly
fourfold, while the S\&P 500 index has gained just 50\%. And without technology, the
benchmark would be flat. In January 2000 alone, 30\% of net inflows into mutual funds
went to science and technology funds, versus just 8.7\% into S\&P 500 index funds.
As a consequence, 
the average price-over-earning ratio P/E for Nasdaq companies is above 200
(corresponding to a ridiculous earning yield of $0.5\%$), a stellar
value above anything that serious economic valuation theory would consider reasonable.
It is worth mentioning that the very same concept and wording of a ``New Economy''
was hot in the minds and mouths of investors in the 1920's. The new technologies
of the time were General Electric, ATT and other electric and communication companies,
and they also exhibited impressive price appreciations of the 
order of hundreds of percent in a 18 month
time intervals before the 1929 crash.

\begin{figure}
\begin{center}
\epsfig{file=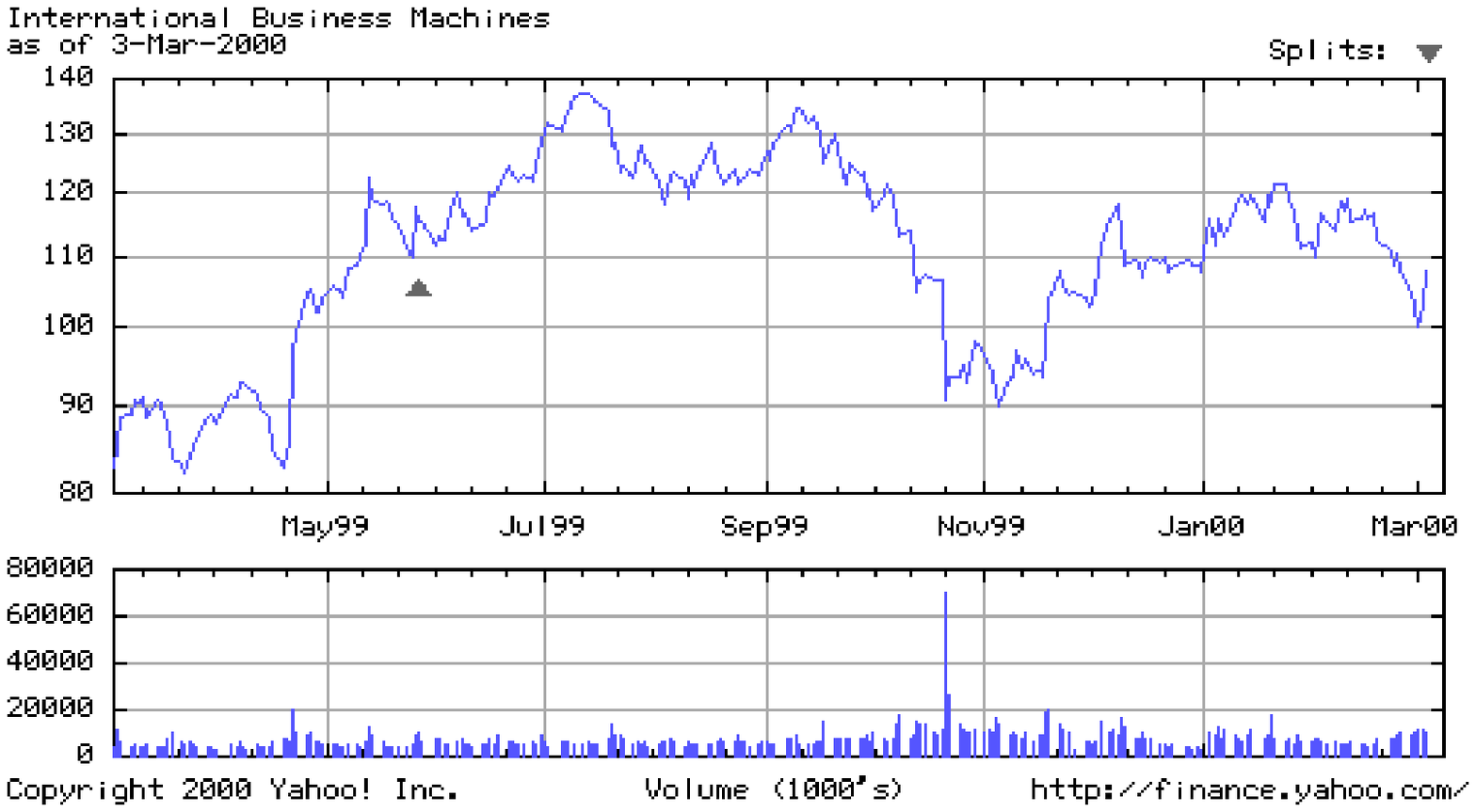,height=8cm,width=18cm}
\caption{\protect\label{IBM} Time series of daily closes and volume of the IBM stock over a
one-year period around the large drop of 21-Oct-1999.  The time of the crash can be seen
clearly as coinciding with the peak in volume. 
Taken from http://quote.yahoo.com/
}
\end{center}
\end{figure}

\begin{figure}
\begin{center}
\epsfig{file=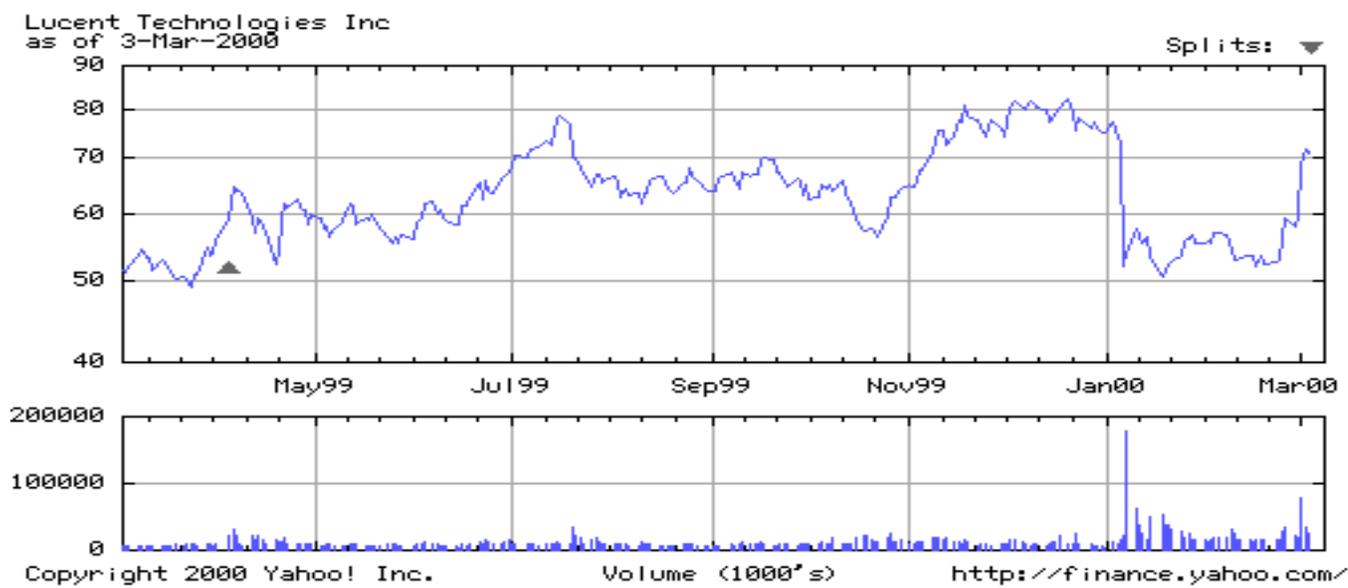,height=8cm,width=18cm}
\caption{\protect\label{LU}  Time series of daily closes and volume of the 
Lucent technology stock over a one-year
period around the large drop of 6-Jan-2000. The time of the crash can be seen
clearly as coinciding with the peak in volume. Taken from http://quote.yahoo.com/
}
\end{center}
\end{figure}

\begin{figure}
\begin{center}
\epsfig{file=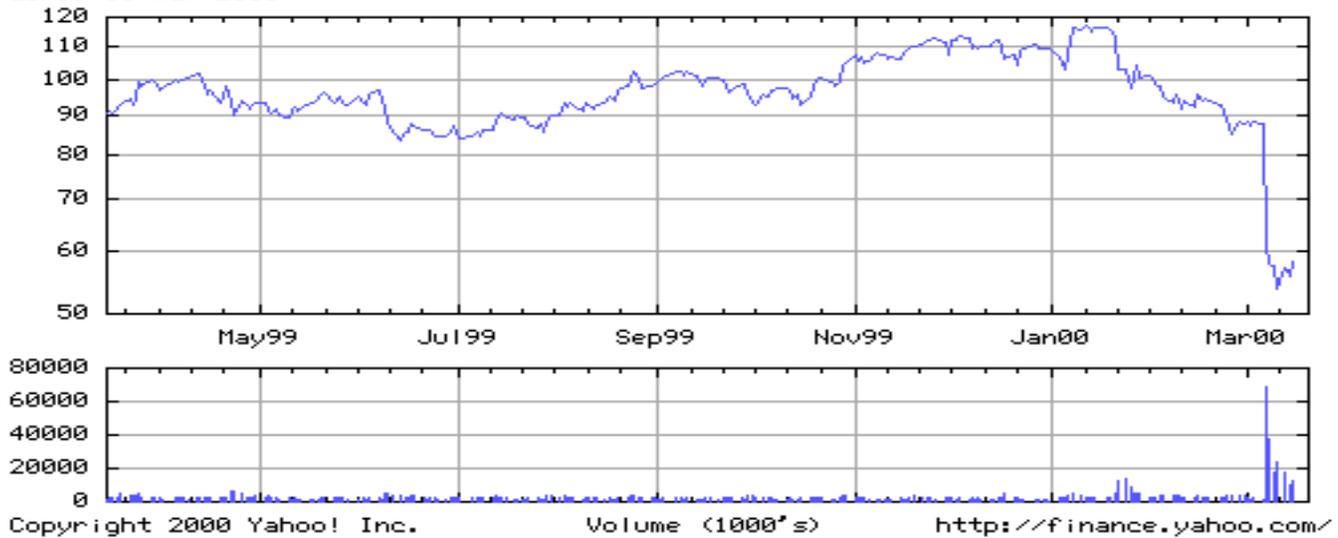,height=8cm,width=18cm}
\caption{\protect\label{PG} Time series of daily closes and volume of the 
Procter \& Gamble stock over a one-year
period ending after the large drop of 7-Mar-2000. The time of the crash can be seen
clearly as coinciding with the peak in volume.
Taken from http://quote.yahoo.com/
}
\end{center}
\end{figure}

Another noteworthy phenomenon is the anomalous occurrence of large sudden losses
(not necessarily always associated with the `new economy' stocks).
For instance, 
three of the ten largest companies in terms of their capitalization on the US market
have recently seen their stock price crash.
IBM (P/E $\approx 30$)
dropped 33\% from a top of 133.31 on 14-Sep-99 to a low of 90.25 on 5-Nov-99 and
during that period 
dropped 15\% from 107 on 20-Oct-1999 to 91 on 21-Oct-1999 in one day (see figure \ref{IBM}), 
due to announcement 
by IBM executives of lower than expected earnings, that surprised Wall Street.
This 33\% drop corresponds to a market capitalization loss of about \$73 billion.
Lucent Technologies (P/E $\approx 100$, P/d $\approx 900$), the telephone equipment giant, 
dropped 28\% from a close of 72.375 on 5-Jan-2000 to a close of 52 on 6-Jan-2000
(see figure \ref{LU})
after the telephone equipment maker warned that its first-quarter
earnings will fall short of expectations, surprising investors.
This 28\% drop corresponds to a market capitalization loss of about \$56 billion.
Procter \& Gamble (P/E $\approx 22$), 
the Cincinnati-based maker of Tide laundry detergent and Crest toothpaste,
 dropped 30\% from 87.44 on 6-Mar-2000 to 61 on 7-Mar-2000
leaving the stock at its lowest since April 1997 (see figure \ref{PG})
after it caught analysts and investors off guard with a profit warning that sent its stock and
those of other consumer products makers in the United States and overseas plummeting.
This 30\% drop corresponds to a market capitalization loss of about \$22 billion.

\begin{figure}
\begin{center}
\epsfig{file=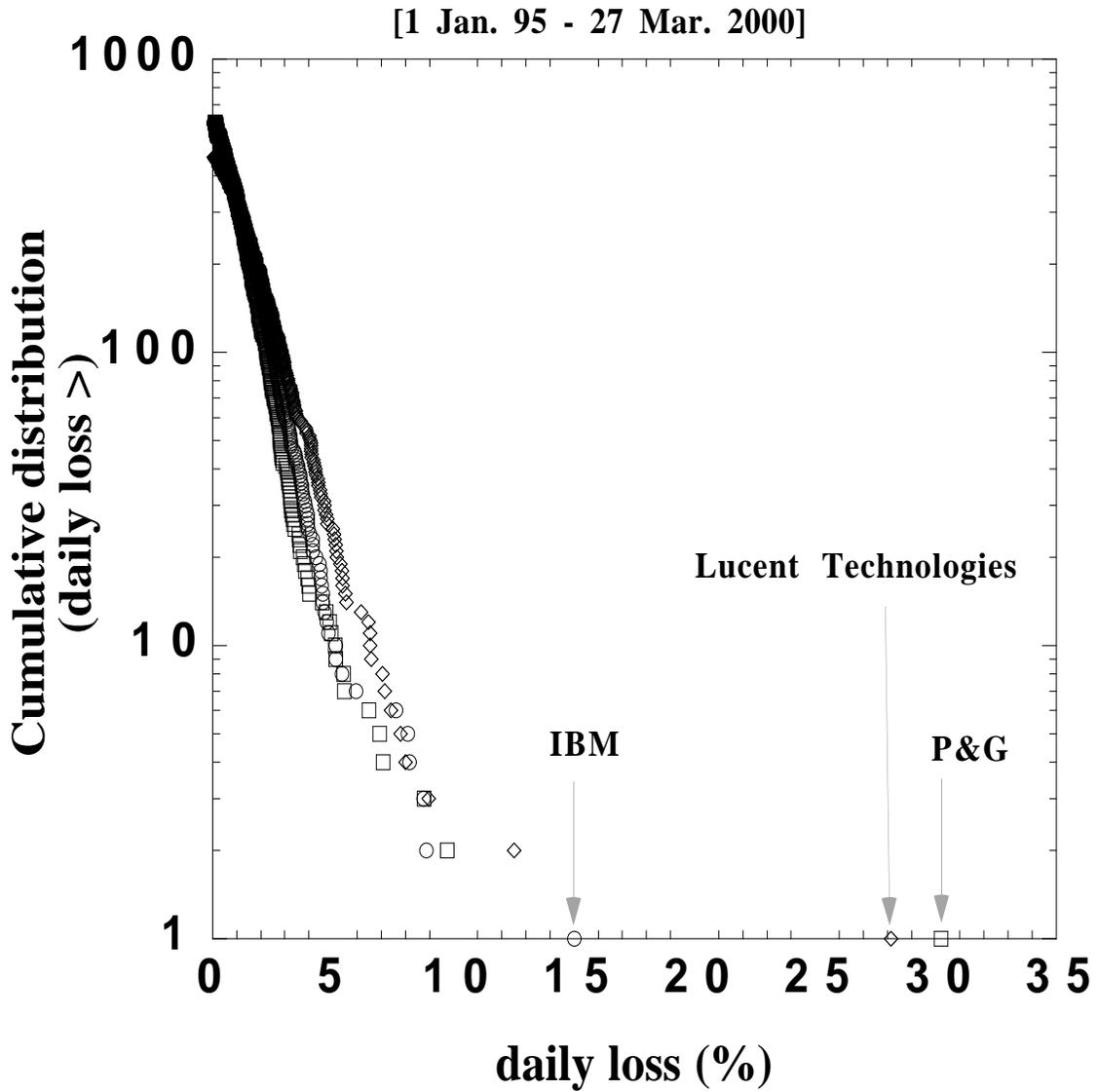,height=15cm,width=15cm}
\caption{\protect\label{FigOutlier} (Complementary) cumulative
distribution of negative daily returns  
$(p(t)-p(t-1))/p(t-1)$ expressed in percent
for the period from 1st January 1995 to 27 March 2000, for the three compagnies 
IBM (circles), Lucent technology (diamonds) and Procter \& Gamble 
(squares) shown in figures \ref{IBM}-\ref{PG}. 
In this semi-logarithmic
representation, a straight line qualifies an exponential distribution. 
Note the position of the three outliers, corresponding to the large drops 
of 21-Oct-1999 for IBM, of 6-Jan-2000 for Lucent technology and of 
7-Mar-2000 for Procter \& Gamble.}
\end{center}
\end{figure}

These enormous events are outliers in the following sense. As seen on figure 
\ref{FigOutlier}, which shows the (complementary) cumulative
distribution of negative daily returns  
$(p(t)-p(t-1))/p(t-1)$ expressed in percent, for the three compagnies 
IBM, Lucent technology and Procter \& Gamble 
for the period from 1st January 1995 to 27 March 2000, most of the negative returns
can be consistently described by an exponential distribution, 
qualified as a straight line in this
semi-logarithmic representation \cite{Lahsor}, with a typical negative return
approximately equal to $3\%$. However, the three recent dramatic daily losses on 
these three companies stand out very clearly off the tail of the exponential 
distributions. Extrapolating the exponentials, this leads to an average repeat
time of about a century for the IBM 15\% drop, and of the order of 
thousands of centuries for the two other events. This confirms the picture found for 
the Dow Jones index 
\cite{outlier} that there exists special times where extraordinary events occur, 
suggestive of some amplification of an instability.

These observations and many others, including so-called speculative bubbles and crashes
\cite{White,JSL}, have led to question the 
idea, cherished by a long tradition of academic researchers,
 that markets are efficient and rational, with alternatives such as
human psychology and behavior in financial decision
making, bounded rationality, agent-based modeling and evolutionary
game theory being actively investigated. 
Here, we advance the novel idea that there is a deep connection between
firm-foundation pricing and speculative phases: they can be seen as two phases
of the same dynamics separated by a bifurcation or ``phase transition''.
If the risk-adjusted expected growth rate is large enough, we find that
a novel speculative-like regime can spontaneously appear which can sustain large price
values. This regime is characterized by a large sensitivity with respect to
expectations, which may explain the observed large reactions of the market.

An important consequence of our proposed theory is that ``speculation''
is not a distinct ``animal'' but rather the ``mutation'' of normal
(approximately) rational behavior.
We do not imply that all speculative behaviors can be fully quantitatively understood
from our approach; rather, we envision the speculative phase emerging from the spontaneous
price symmetry breaking as the nucleus for speculation blossoming with imitation
and behaviorial processes.

In section 2, we
summarize the firm-foundation theory and discuss its consequences.
In section 3, we present the concept of price parity symmetry
and show how, together with the 
firm-foundation theory, it leads to the concept of a critical bifurcation
to a spontaneous symmetry breaking regime
and suggest the existence of a ``speculative phase'' linked to the fundamental
valuation phase. Section 4 proposes a simple
dynamical extension of the firm-foundation theory that captures both regimes and leads
to observable predictions. Section 5 analyses finite time-horizon effects and 
the effect of future prices on expectation of growth. Section 6 concludes.

\section{Equilibrium economic valuation: effect of expected dividend growth}

The ``firm-foundation'' theory asserts
that a stock has an intrinsic value determined by careful analysis of present conditions
and future prospects. Developed by S. Eliot Guild \cite{Guild} and John B. Williams
\cite{Williams}, it is based
on the concept of discounting future dividend incomes. In the words of Burton G. Malkiel
\cite{Malkiel},
discounting refers to the following concept:
``rather than seeing how much money you will have next year (say \$1.05 if your put
\$1 in a saving bank at 5\% interest), you look at money expected in the future and
see how much less it is currently worth (thus next year's 1\$ is worth today only
about 95c, which would be invested at 5\% to produce \$1 at that time).''
In practice, the intrinsic value approach is
a quite reasonable idea which is however confronted with slippery estimations:
the investor has to estimate future dividends, their 
long-term growth rates as well as the time-horizon over which the growth rate will be maintained.
Notwithstanding these problems, this approach has been promoted by Irving Fisher \cite{Ifisher}
and Graham and Dodd \cite{GrahamDodd} so that generations of Wall 
Street security analysists have been using
some kind of ``firm-foundation'' valuation to pick their stocks.

Consider a company paying a dividend $d$ per share at regular time intervals $\tau$, $2 \tau$, ...
Let $r$ be the fixed risk-free interest rate (the ``price of money'')
paid by investing one unit of wealth over a time $\tau$.
The time discount factor per unit time $\tau$ 
is thus $1/(1+r)$: one unit of weath at time $t=\tau$ is worth $1/(1+r)$ at present $t=0$.

In the firm-foundation equilibrium economic theory \cite{Debreu}, 
the equilibrium price $p$ of a share is the 
present expected discounted value of all future payoffs:
\be
p = d + {d \over 1+r} + {d \over (1+r)^2} + ... = (1+r) {d \over r}~. \label{hfkjks}
\ee
This standard result expresses a no-arbitrage condition as the instantaneous
return $d/p$ obtained by owning the stock is essentially equal to the risk-free rate.
The price-over-dividend ratio $p/d$ is 
\be
 {p \over d} = {1+r \over r}~. \label{jfka}
\ee
The price-over-earning ratio uses the company's earning (total revenues minus expenses after tax).
The price-over-dividend ratio is function of the company's strategy of 
how much of the earnings should be
distributed to the share holders in the form of dividends rather than reinvested in the company
or used to pay debts.
Earnings and dividends are empirically 
correlated and exhibit parallel evolutions but they have different explanatory and
predictive powers of price returns. Lamont \cite{Lamont} has shown 
recently that the price-over-dividend
ratio forecasts excess returns of stocks: high dividends forecast high returns. In contrast,
high earnings forecast low returns. One interpretation is that dividends measure the 
permanent component of stock prices, due to managerial behavior in setting dividends because
they help measure the future dividends. The level of earnings is a good measure of current
business conditions: a more favorable economy implies larger earnings, for which investors
require lower future returns as risk premia, hence larger earnings predict smaller returns.

Let us now consider the possibility for the dividend to grow as a function of time
at the fixed rate $r_d$, as a result of increased productivity and other factors. $r_d$ has
to be understood as the expected growth rate of future dividends. Usually, growth 
forecasts are performed on the company's growth, in terms of its earnings.
We shall assume for simplicity that both dividends and earnings have 
statistically the same growth rate.
Equation (\ref{hfkjks}) is then changed into
\be
p = d + d{1+r_d \over 1+r} + d{ (1+r_d)^2 \over (1+r)^2} + ... = (1+r) {d \over r-r_d}~. 
\label{hfkaajkas}
\ee
The price-over-dividend ratio is then changed into
\be
{p \over d} = {1+r \over r-r_d}~.  \label{jfkakka}
\ee
This is the celebrated valuation formula of Gordon and Shapiro \cite{Gorshap}.

Expression (\ref{jfkakka}) accounts for the impact on the stock price
of the expected dividend payout $d$, 
the growth rate $r_d$ and the market interest rate $r$. The last important ingredient
is the degree of risk. There is no unique measure of risk. For our purpose, 
risk embodies several components, such as market price and earning
volatilities as well as growth rate uncertainty and variability. 
To quantify the impact of these risks on
the value $p$ of a stock, we
shall adopt the pragmatic procedure in which the growth rate $r_d$
is adjusted for risk by replacing it by $r'_d=r_d - R$, where $R$ is an aggregate risk level. 
In the spirit of a quality ratio $r_d/R$, such as the Sharpe ratio, $R$ 
renormalizes $r_d$. Errors in estimating $R$ will increase the uncertainties 
on $r'_d$ already present due to the large uncertainties in the growth rate $r_d$.
Expression (\ref{jfkakka}) is thus replaced by
\be
{p \over d} = 1 + {1+r_d -R \over 1+r} + {(1+r_d -R)^2 \over (1+r)^2} 
+ ... = {1+r \over r+R-r_d}~.  \label{jfkaaaakka}
\ee

In accordance with intuition, the price-over-dividend increases with dividend and
with the rise of 
expected dividend due to a positive growth rate, i.e. with the expectation that future gains will 
grow.
This expression (\ref{jfkaaaakka}) reveals interesting perhaps less intuitive phenomena.
When the risk-adjusted dividend growth rate $r'_d=r_d-R$ 
is not small compared to the risk-free interest rate $r$, 
(\ref{jfkaaaakka}) shows that the price $p$ can be many times larger than the valuation 
formula (\ref{hfkjks}) would lead us to believe. Take for instance, $r=5\%$, 
 $r+R=15\%$ and $r_d=14.5\%$.
Expressions (\ref{jfka}) and (\ref{jfkaaaakka}) give respectively ${\rm P/d} = 21$ and $210$.
In this framework, large price-over-dividend ratios express the large 
expected growth rate of future earnings. This valuation formula (\ref{jfkaaaakka}) may 
actually form the basis for a partial
rationalization of some large observed price-over-dividends
and probably also price-over-earning ratios. This is
particularly true for the ``new economy'' based on high-growth information- and 
internet-based companies that are expected to develop large future earnings. 
What (\ref{jfkaaaakka}) teaches us is that their pricing may actually be not as much irrational
as often believed
but rather reflects the high future growth of the expected earnings. The new emphasis
is not so much on the present dividend $d$ but on its future risk-adjusted growth rate $r'_d$.
This view has been expounded by many workers (see intuitive presentations in
 \cite{Malkiel} and \cite{Zajbook}). 
 
What happens when the risk-ajusted growth rate $r'_d$ becomes larger than $r$, so that
the valuation formula (\ref{jfkaaaakka}) gives a meaningless negative price?
In the economic literature, this regime is known as the growth stock paradox
\cite{Blanchfish}. This is the question addressed in the sequel.

Note that the valuation problem has been posed in 1938 by Von Neumann 
\cite{vonneumann} who demonstrated that, in an economy with balanced growth,
the growth rate is always identical to the interest rate and thus equal to the
discounted rate. Zajdenweber \cite{Zajbook} then points out that the value of a share
is, as a consequence, always infinite as seen from (\ref{hfkaajkas}). 
The intuition is that when $r'_d$ becomes
equal to (and this is all the more true when it is larger than) $r$, the price
of money is not enough to stabilize the economy: it becomes favorable to borrow money
to buy shares and earn an effective rate of return, which is positive
for all values of the dividend. This is exactly what happened on the US market 
in the rally preceeding the Oct. 1929 crash \cite{Galbraith}.
Note that a negative $r-r'_d$ is similar to a negative interest
rate $r$ in absence of growth and risks: it leads to an arbitrage opportunity since
you can borrow \$1 now, keep it under your mattress,
and give back $\$1 \times (1-|r|)$ at a later time, pocketing $100 |r|$ cents in the process.
Another well-known situation of negative $r-r'_d$ 
is when the inflation rate is larger than the interest rate. Tax laws which tax the 
nominal income and ignore inflation reduce the effective $r$ even more.
Here again, you lose money by keeping it.

The valuation formula (\ref{hfkaajkas}) can be easily generalized to account for
arbitrary time-varying dividends and rates
\be
P/d = 1 + a_1  + a_1 a_2 + ... + a_1 a_2... a_n  + ...  ~,  \label{jhhpqnvzz}
\ee
where $a_n$ is the global ``discount'' factor $(1+r_d-R)/(1+r)$ at year $n$ in the future,
taking into account a possible variation of the risk-free interest rate and of the 
risk-adjusted dividend
growth rate (which can be negative). The sum of products
given in the r.h.s. of (\ref{jhhpqnvzz})
has been studied in a variety of contexts (see \cite{kestenreview} for a review) including
a mechanism for Pareto distributions of cities 
\cite{Champenowne} and in relation with ARCH(1) processes \cite{Haan}.
The asymptotic behavior of expression (\ref{jhhpqnvzz}) is well-understood: 
\begin{itemize}
\item for 
$\langle \ln a \rangle < 0$, where the brackets denote the expectation with respect to the 
distribution of $a_n$, $p/d$ is finite and is distributed with a power law tail with exponent
depending sensitively on the distribution of the $a_n$'s \cite{kestenreview,Sorcont}. This
is reminiscent of the fat tail character of price returns (see \cite{luxsor} for the relation 
between prices and price returns in this context) and of the sensitivity of prices
to external news that may be seen to impact
 the estimation of the risk-adjusted growth rate $r'_d$.

\item For $\langle \ln a \rangle > 0$, the sum grows on average exponentially fast with the number of
terms (time horizon) with an average growth rate equal to $\langle \ln a \rangle$, while
exhibiting large fluctuations. This regime is qualitatively the same as that seen
in (\ref{jfkaaaakka}) for ${1+r_d -R \over 1+r} > 1$, i.e.  $r < r_d -R$.

\item The critical transition point $\langle \ln a \rangle = 0$
is in general also characterized by a growth as the number of terms included increases, but 
this growth is sub-exponential and depends on details of the distribution and on correlations
\cite{calanetal}. 
\end{itemize}

\section{Spontaneous price valuation}

\subsection{Negative prices and price parity symmetry}

Expressions (\ref{hfkaajkas}) and (\ref{jfkaaaakka}) have no meaning for $r'_d > r$, as the
price become negative which makes no apparent sense. This argument that prices should
always remain positive was actually at the 
basis of the rejection of Bachelier's seminal random walk model of prices \cite{Bachelier}
in favor of Samuelson's 
random walk model of the logarithm of prices \cite{Samuelson}, which has become the 
paradigm in economic and finance modeling.

Actually, it makes perfect sense to think of {\it negative} prices. We are ready to pay a 
(positive) price for a commodity that we need or like. However, we will not pay a positive
price to get something we dislike or which disturb us, such as garbage, waste,
broken and useless car, chemical and industrial hazards, etc.
Consider a chunk of waste. We will be ready to
buy it for a negative price, in order words, we are ready to take
the unwanted commodity if it comes with cash. This exchange of waste for income
is the basis for the industry of waste management. Nuclear waste from some countries
such as Japan for instance are shipped to La Hague reprocessing complex
 in France, which is ready to store
the unwanted wastes for income. The Japanese are thus paying a price to get rid 
of their waste, that is to say, La Hague is paying a negative price to get the 
nuclear waste commodity! As a matter of fact, this exchange of wastes 
is at the basis of a huge business
for the present and future management of industrial and nuclear waste that
counts in hundreds of billions of dollars. A less obvious example is
the case of electricity companies which sell surplus electricity
in exceptional cases for negative prices; it is expensive for them to shut down 
a power plant and to restart it again (R. Weron, private communication).
Stauffer humoristically
points out (private communication) that the page charges some authors pay 
to the journals to get rid 
of their manuscripts are a nice example of negative prices. Actually, this is
not correct but this example
illustrates the intricacy of the concept: authors pay
to get published, not get rid of their paper but to buy fame, i.e. cash leaves the authors
but fame comes to them (hopefully), hence the positivity of the price.

\begin{figure}
\begin{center}
\epsfig{file=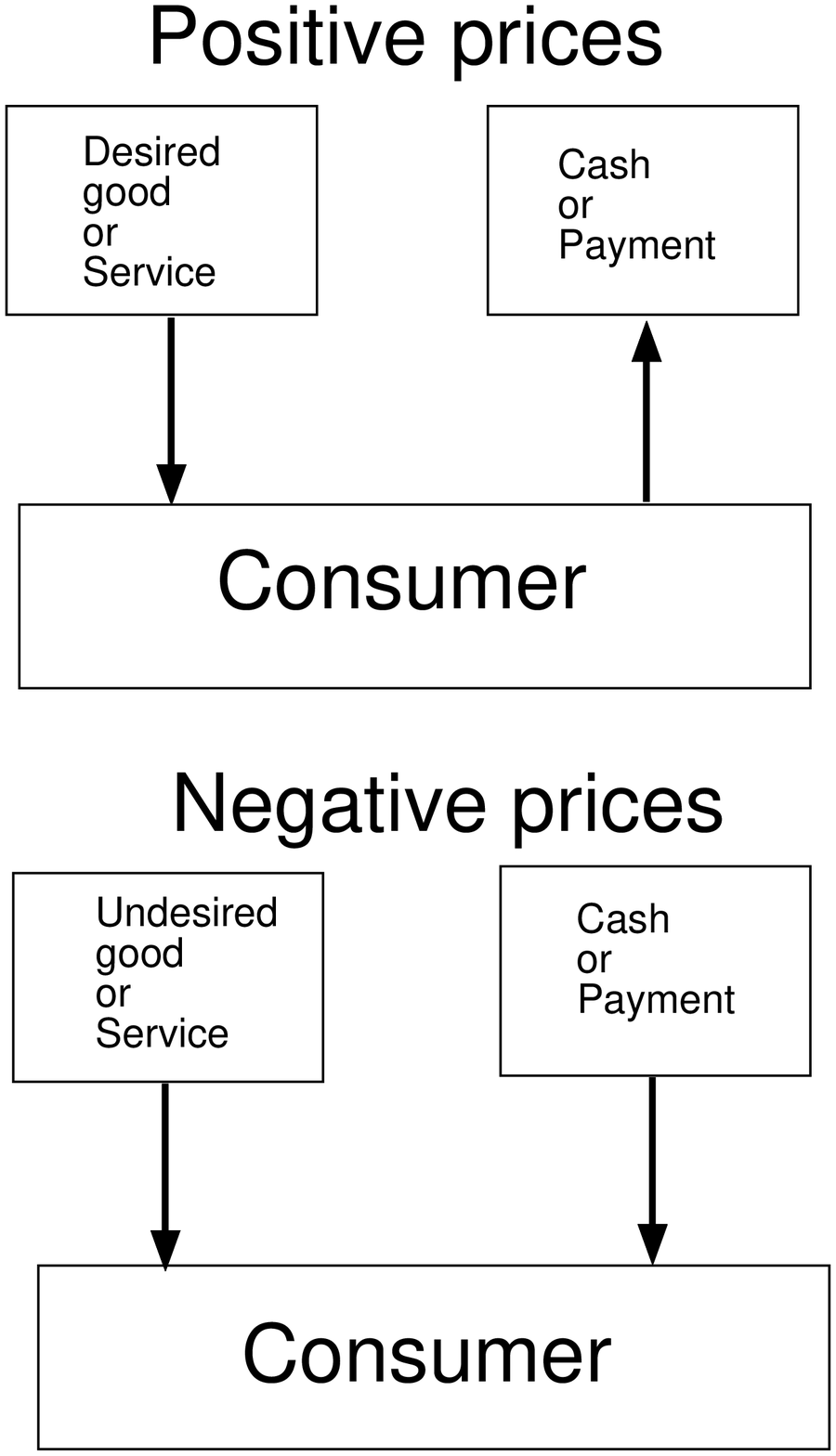,height=18cm,width=15cm}
\caption{\protect\label{Posnegprice} Cartoon showing that the sign of price
is defined by the relative direction of the flux of cash or payment compared
to the flux of good or service: a positive price corresponds to the more
commonly experienced situation where the cash or payment flux is with 
a direction opposite to the flux of good or service; a negative price corresponds
to the reserve situation where the cash or payment flux has the same
direction as the flux of good or service.
}
\end{center}
\end{figure}

In sum, we pay a positive price for something we like and a negative price for
something we would rather be spared of (i.e. we pay a (positive) price to get rid of it or 
we need a remuneration to accept this unwanted commodity). This concept is illustrated in
figure \ref{Posnegprice}.

In the economy, what makes a share of a company desirable? Answer: its earnings,
that provide dividends, and its potential appreciation that give rise to
capital gains. As a consequence, in absence of dividends and of speculation, 
the price of share must be nil. There is thus a natural 
\be
p \to -p ~~~~{\rm parity ~symmetry}~,    \label{fjallafc}
\ee
where both
positive and negative prices quantify our liking or disliking of the commodity. 
The earnings leading to dividends $d$ thus act as a symmetry-breaking ``field'',
since a positive $d$ makes the share desirable and thus develop a positive price. 
It is clear that a negative dividend, a premium that must be paid regularly 
to own the share, leads to a negative price, i.e. to the desire to get rid of that stock
if it does not provide other benefits. Andersen et al. \cite{technianaly}
have also used the price, rather than its logarithm, to develop a fundamental framework
of price patterns. 

For a share of a company that is neither providing utility nor a waste,
there is no intrinsic
value for it if it does not give you more buying power for something you desire. 
Hence, its price is $p=0$ for $d=0$. We can allow for both 
positive and negative price fluctuations, but there is a priori nothing that 
breaks the symmetry (\ref{fjallafc}).

We stress that the price symmetry (\ref{fjallafc}) is distinct from the 
gain/loss symmetry of stock holders, before the advent of limited liability companies
in the middle of last century. With the present limited liability of stock holders,
owning a stock is akin to hold an option: gain is accrued from dividend and capital gains;
on the downside, losses are limited at the buying price of the stock. This asymmetry,
which is a relatively recent phenomenon and led to the full development of capitalism,
is conceptually distinct from the symmetry (\ref{fjallafc}) of prices.

These considerations can be summarized by
the following analogy between valuation and physical phase
transitions (liquid-gas or paramagnetic-ferromagnetic)
 as, for instance, examplified by the canonical Ising model \cite{Staubook}:
\bea
{\rm price} ~p  & \Longleftrightarrow &  {\rm order ~parameter} \nonumber\\
{\rm riskless~ interested~ rate~ above ~ risk-adjusted~
dividend~growth~rate}~ r-r'_d  & \Longleftrightarrow & {\rm 
reduced~control~ parameter}  \nonumber\\
{p \over d} & \Longleftrightarrow & {\rm susceptibility}~\chi  \nonumber\\
{\rm dividend}~d & \Longleftrightarrow & {\rm external~ field}~H~.  \nonumber
\eea
In this analogy, the price equation (\ref{hfkaajkas}) becomes
\be
p = \chi H~.   \label{jfjlala}
\ee
Notice again that the positive value of the price is associated with the positivity of the dividend.
Expression (\ref{jfjlala}) takes the standard form for the dependence of the order parameter
as a function of the external field $H$ within linear response theory \cite{Goldenfeld}. 
The coefficient
of proportionality $\chi$ is seen to diverge as the control parameter $r-r'_d$ approaches
$0$
\be
\chi \sim (r-r'_d)^{-\gamma}~, ~~~{\rm with~the~standard~mean~field~value~} \gamma=1~.
\label{jfhjamama}
\ee

\subsection{Speculation and spontaneous price symmetry breaking}

This analogy suggests a very natural interpretation of the regime where
the risk-ajusted dividend growth rate $r'_d$ becomes larger than the risk-free rate $r$:
for negative reduced control parameter, the price $p$ can become non-zero
even {\it in absence} of dividend by a mechanism known as ``spontaneous 
symmetry breaking'' (SSB).

\begin{figure}
\begin{center}
\epsfig{file=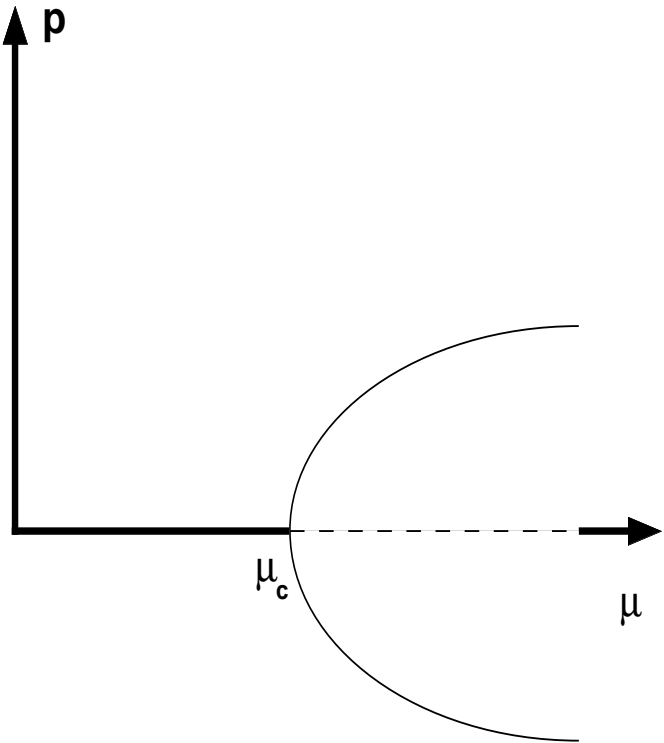,height=10cm,width=10cm}
\caption{\protect\label{FigSuperbifurcdraw} Bifurcation diagram, near
the threshold $\mu_c$, of a ``supercritical'' bifurcation. The order parameter 
$p$ bifurcates from the symmetrical state $0$ to a non-zero value $\pm p_s(\mu)$
represented by the two branches, as the control parameter crosses the critical
value $\mu_c$. The value $p=0$ represented by the dashed line becomes unstable
for $\mu > \mu_c$. 
}
\end{center}
\end{figure}

SSB is one of the most important concept in modern science that underpins 
our present understanding of the universe and of its interactions. 
Its basic principle can be illustrated by a very simple dynamical system
\be
{dp \over dt} = (\mu - \mu_c) p - {p^3 \over p_s^2}~.
\label{eqzee} 
\ee
This equation possesses the parity symmetry (\ref{fjallafc}), since both
$p$ and $-p$ are solution of the same equation. A solution respecting this symmetry
obeys the symmetry condition $p=-p$ whose unique 
solution $p=0$ is called the symmetry-conserving solution.
For $\mu < \mu_c$, $p$ is attracted to zero and the 
asymptotic solution $p(t \to +\infty)$ is zero, which as we said is the only solution
respecting the parity symmetry. 
However, a solution of (\ref{eqzee}) may not always
respect the parity symmetry of its equation. 
This occurs for $\mu > \mu_c$ for which the asymptotic solution is
\be
p(t \to +\infty) = \pm p_s~ (\mu - \mu_c)^{1 \over 2}~,~
\label{esgdg}
\ee
where $p_s$ gives the characteristic scale of the amplitude of p. 
Notice that there are now two distinct solutions, each of them being 
related to the other by the action of the parity transformation $p \to -p$:
the set of solutions respects the parity symmetry as an ensemble but each
solution separately does not respect this symmetry. This phenomenon is
called ``spontaneous symmetry breaking'' (SSB). More generally, 
the concept of SSB describes the situation in which a solution has a lower
symmetry than its equation. Expression (\ref{eqzee}) shows that 
the parameter $\mu$ controls the symmetry breaking. The so-called
``super-critical bifurcation'' diagram near
the threshold $\mu = \mu_c$
synthetizing the transition from a symmetric solution $p=0$ to a SSB solution
is shown in figure \ref{FigSuperbifurcdraw}. 

\begin{figure}
\begin{center}
\epsfig{file=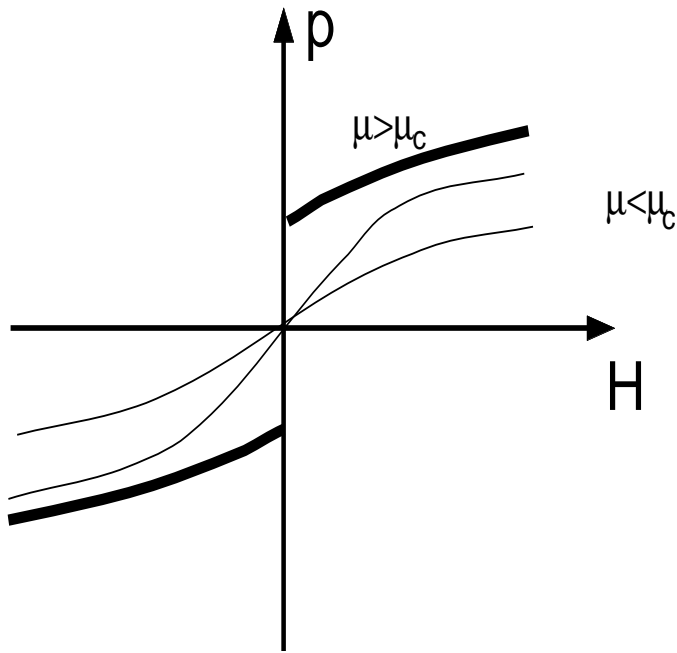,height=10cm,width=10cm}
\caption{\protect\label{FigJumpOPdraw} Order parameter $p$
as a function of the external field for different values of the 
control parameter $\mu$. The two thin lines correspond to two different
values of $\mu < \mu_c$. The thick line is the spontaneous symmetry broken phase
occurring for $\mu > \mu_c$.
}
\end{center}
\end{figure}

The SSB concept takes its full meaning in presence of a very small external ``field''
$H$ such that (\ref{eqzee}) is transformed into
\be
{dp \over dt} = H + (\mu - \mu_c) p - {p^3 \over p_s^2}~.
\label{eqzeaae} 
\ee
For $\mu < \mu_c$, the stationary solution of (\ref{eqzeaae}) is given by
(\ref{jfjlala}). In constrast, in the SSB regime $\mu > \mu_c$, $p$ jumps
from $ p_s~ (\mu - \mu_c)^{1 \over 2}$ to $- p_s~ (\mu - \mu_c)^{1 \over 2}$
when $H$ goes from positive to negative as illustrated in figure \ref{FigJumpOPdraw}:
any infinitesimal field is enough to flip abruptly the order parameter $p$ from
one of its two symmetry-broken solutions to the other. We refer to \cite{SSBrefs} for
general references and pedagogical introductions and examples. 
It will never be sufficiently stressed how important is this concept of SSB:
for instance, it is invoked to unifying fundamental interactions: weak, strong and 
electromagnetic interactions are now understood as the result of a more fundamental
spontaneous symmetry broken interaction \cite{Weinberg}. 
In another sweeping application, particles and matter in this universe seem
to be the spontaneous symmetry-broken phases of a fundamental vacuum state \cite{Weinberg},
similarly to the non-vanishing price emerging in the SSB phase $\mu > \mu_c$
out of the symmetry-conserved ``vacuum'' solution $p=0$.
Critical phase transitions are also understood as SSB phenomena \cite{Goldenfeld}.

To come back to the valuation problem, we thus propose that, for $r < r'_d$,
assets acquire a spontaneous valuation as a result of this spontaneous symmetry breaking
mechanism.
When $r-r'_d$ become negative, money is not a desirable commodity. You lose money by keeping it.
Other commodities become valuable in comparison with money, hence, the spontaneous 
price valuation in absence of dividend. 
We thus propose that, for $r-r'_d < 0$,
the price becomes spontaneously positive (or possibly negative depending on initial
conditions or external constraints) and this spontaneous valuation is nothing but the appearance of 
a {\it speculation} regime: investors do not look at or care for dividends;
the increase of price is self-fulfilling.
And the interpretation of speculation (defined
as price higher than fundamental $=0$ here) is a spontaneous symmetry-breaking around
the value $p=0$.
              
According to our theory, the regime $r<r'_d$ is a self-sustained growth regime where
prices become unrelated to earnings and dividends: prices can go up independently of the
dividends due to the ``spontaneous symmetry breaking'' where a company's share acquires
spontaneously value without any earning. This situation is similar to the spontaneous
magnetization of iron at sufficiently low temperature which acquires a spontaneous
magnetization under zero magnetic field. This regime could be relevant to understand
regimes of bubbles as well as regimes as can be seen in the present
``new economy'' where prices increase resulting in
high price-over-dividend without apparent economic rationalization.

\section{A simple extension of pricing theory to the critical and spontaneous valuation regime}

The question we now address is how to determine the price
at and below the ``critical point'' $r=r'_d$ for
which the standard equilibrium economic approach fails. 

In order to make further progress in the quantification of these different phases,
we propose first to cure the nonsensical valuation expression (\ref{hfkaajkas})
by extending the description into a dynamical framework.
Building on the price symmetry (\ref{fjallafc}), it is natural to consider 
a Taylor expansion around the neutral value $p=0$ to obtain
the simplest possible dynamical equation for price evolution that 
embodies these previous elements. The following Langevin
equation has the same structure as (\ref{esgdg})
\be
T {dp \over dt} = d - {r-r'_d \over 1+r} p -  {p^3 \over p_0^2} +\eta~,   \label{jjkala}
\ee
where $p_0$ is a numerical coefficient with ``dimension'' of price. 
Expression (\ref{jjkala}) embodies four
contributions. The first term $d$ in the r.h.s. is simply stating that
price is driven up by earning. A comparison with (\ref{eqzeaae}) shows that $d$ indeed
plays the role of the external field breaking the symmetry $p \to -p$.
The second term $- {r-r'_d \over 1+r} p$
is limiting this appreciation by the discounting process such that the price would
converge at long times to the equilibrium value (\ref{hfkaajkas}). Note that it
 correspond to a growth of the price only if $r'_d - r$ is positive, i.e. the 
 growth rate is larger than the discount rate.
The third term
$-  p^3/p_0^2$ is such that it is negligible as long as the price is not too large:
$p_0$ quantifies what is meant by ``large.'' When the price become
very large as for instance when $r\leq r'_d$, its growth becomes limited by nonlinear
feedback effects. We choose a cubic term as the first relevant non-linear term,
as it is the first one keeping the symmetry $p \to -p$. 
In the spirit of a Taylor expansion in powers of $p$, expression (\ref{jjkala})
thus contains a constant term (which, as already said, breaks the $p \to -p$ symmetry), 
a linear and a cubic term. Higher order terms
can easily be incorporated but are less important. The fourth term in the r.h.s. of 
(\ref{jjkala}) is a noise with covariance $\langle \eta (t) \eta(t') \rangle = D \delta(t-t')$,
which accounts for the stochastic structure of pricing on stock market.
The constant $T$ in the l.h.s. is a time scale characterizing the reaction time
of the price with respect to a change of any of the terms in the r.h.s. It is typically 
of the order of seconds in liquid markets. In the physical literature, equation (\ref{jjkala})
is well-known and has been studied in different contexts, for instance as a noisy mean-field
time-dependent Landau theory for critical phenomena
or as the noisy normal form of supercritical bifurcations in dynamical
system theory.

Equation (\ref{jjkala}) has not been derived on the basis of direct economic reasoning
but uses instead 1) symmetry arguments and 2) the constraint that it recovers the 
firm-foundation valuation formula for $r > r'_d$.
Direct derivations from standard
valuation settings give different equations where the constant term $d$ is 
usually given a negative sign to account for the fact that, at equilibrium, the value
of the stock is not changed by the dividend payment $p(t^+) = p(t^-) - d$,
where $t$ is the time of dividend payment and the superscripts refer to the time just 
before and after the dividend payment. It is important to realize that 
expression (\ref{jjkala}) accounts
for both economic valuation due to growth and for the impact of dividend and growth rate
on the psychological appreciation of the value of the stock by investors. In other words, 
the symmetry-based equation (\ref{jjkala}) somehow automatically embodies a mixture
of economic and psychological ingredients. This is indeed necessary for this equation to
describe {\it both} the standard firm-foundation and the speculative regimes.
Deriving (\ref{jjkala}) from first principles will require to include a blend of economic and
psychological factors. We hope to report on this in a future communication.
Similarly to phase transitions in the physical sciences,
we realize here again the power of symmetry-derived effective equations, in contrast
to microscopically-derived equations which are much more prone to miss important
contributions. For instance, we have seen that a purely economically-based
valuation equation leads to the growth stock paradox
\cite{Blanchfish}, which cannot be solved without enlarging the theory as we do with 
(\ref{jjkala}).

\begin{figure}
\begin{center}
\epsfig{file=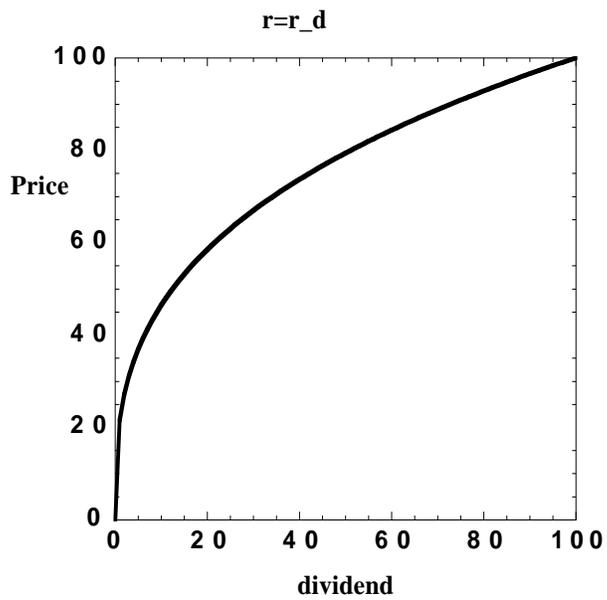,height=8cm,width=8cm}
\caption{\protect\label{Pricedividenc} Dependence of the price 
$p=(p_0^2 d)^{1/3}$ as given by (\ref{jfjaka}) as a function of dividend $d$, using
$p_0=100$.
Observe the very fast rise of the price as the dividend increases from zero, compared
to a linear law. The relevant region of the curve is typically for $d < 10$.
}
\end{center}
\end{figure}

The consequences of equation (\ref{jjkala}) are the following.
At $r=r'_d$, the apparent divergence of the price given by (\ref{hfkaajkas})
is replaced by a non-analytic dependence of the equilibrium price as a function of dividend
(neglecting the noise term):
\be
p \sim d^{1/\delta}~, ~~~{\rm where}~\delta = 3~ ~~~{\rm at}~~ r=r_d~.  \label{jfjaka}
\ee
This one-third power law dependence of the price thus predicts a much faster increase of the price
as a function of earnings than the usual linear dependence (\ref{hfkaajkas})
for small $d$'s as illustrated in figure \ref{Pricedividenc}. 
Indeed, expression (\ref{jfjaka}) has an infinite slope
for $d \to 0$. Expression (\ref{jfjaka}) is a characteristic signature of the critical point
$r=r'_d$. Close to the critical point, changes in expected dividends thus lead to large
price movements. This has recently been seen on IBM, Lucents and Procter \& Gamble,
discussed in the introduction.

We can retrieve (\ref{jfjaka}) by assuming 
a co-integration of $d$ and $r-r_d$ parameterized by
\be
r-r'_d = c \left({d \over p_0}\right)^{\alpha}~,~~~~{\rm where}~0\leq \alpha \leq 1~,
\label{ncbbzvvz}
\ee
valid for small $d$ and $r-r'_d$. This expression implies that, as $r-r'_d \to 0$,
the price can remain well-behaved if $d$ also goes to zero with the appropriate relationship.
This is seen by plugging the expression (\ref{ncbbzvvz}) into (\ref{hfkaajkas}) which yields
\be
p = (1+r) p_0^{\alpha}~d^{1-\alpha}~.
\ee
This retrieves the conceptual content of (\ref{jfjaka}) and quantitatively gets the same
exponent
with the choice $\alpha = 2/3$.

For $r<r'_d$ such that the expected risk-adjusted growth of earnings is faster that the riskless
interest rate, the prices tend to grow spontaneously even without visible earning ($d \sim 0$).
This is the regime of ``spontaneous valuation'', similar to the ``spontaneous symmetry
breaking'' discussed in the previous section.
This may explain the present frenzy on ``growth companies'' which have apparently 
ridiculously high price-over-dividend ratios. 
In this case, we see that the equilibrium price is set by 
\be
p \propto p_0 (r'_d - r)^{1/2}~,     \label{jfllbala}
\ee
independently of earnings $d$ for small $d$'s as seen in figure \ref{FigJumpOPdraw}. 
The rational for paying a high price
without earning is that any possible future non-zero earnings will grow sufficiently fast
so that it will become valuable. This is the regime of speculation.

The general stationary solution of (\ref{jjkala})
that combines the two regimes (\ref{jfjaka}) and (\ref{jfllbala})
takes the form \cite{Goldenfeld}
\be
p = p_0 (r'_d - r)^{1/2}~ f\left( {d/p_0 \over  (r'_d - r)^{\delta/2}} \right)~,  
\label{fjjlalaZ}
\ee
where the function $f(x) \approx 1$ for small $x$ and 
$f(x) = x^{1/\delta}$ for large $x$. The first condition ensures that (\ref{fjjlalaZ})
retrieves  (\ref{jfllbala}) for small or vanishing dividends. The second condition
allows to retrieve  (\ref{jfjaka})  at or very close to the critical point. The specific
functional form of the cross-over of $f(x)$ at intermediate values of $x$ depends on higher
order terms that have been dropped out of this description. 

The noise term $\eta$ in (\ref{jjkala}) ensures that the price $p(t)$ is stochastic.
The correlation of price returns can be explicitely calculated and takes the form
\be
\langle {\rm return}(t) {\rm return}(t')\rangle =
{D (1+r) \over 2 |r-r'_d|} \exp \left[-{|t-t'| \over \tau}\right]~,
\ee
where the correlation time is
\be 
\tau = T {1+r \over  |r-r'_d|}~.    \label{fjaakla}
\ee
In agreement with empirical observations, returns are correlated only over
a very short time scale proportional to $T$ (since we have neglected the effect of
transaction costs, $T$ is of the order of the typical time for a trade, i.e. seconds), 
except when $r$ approaches
$r'_d$ for which the correlation time can be longer. The mathematical divergence
of the correlation time (\ref{fjaakla}) is very difficult to observe in practice
because the amplification factor $1/|r-r'_d|$ really kicks off $\tau$ when 
$|r-r'_d|$ is quite close to $0$. If $T=10$ seconds, say, and $|r-r'_d|=5\%$, 
(\ref{fjaakla}) gives $\tau \approx 3$ minutes. $|r-r'_d|$ needs to be of the order
of $1\%$ or smaller for the correlation time to be of the order of $15$ minutes or larger.
This suggests the following interpretation: when investors estimate the risk-ajusted 
growth rate in the range of the risk-free interest rate, pockets of predictability
can be created. Inversely, it is sometimes argued by operators and managers of funds
that the market exhibits ``pockets of predictability''. Such non-stationary structures
could be attributed to a particular risk and growth estimation by investors collectively
at some particular times.

\section{Market instabilities and finite time horizons}

\subsection{Amplification of price variations by correlations between dividend and
its growth rate}

The previous framework allows us to cast some light on apparent anomalies and on
surprising market behavior. 
According to (\ref{hfkjks}), the announcement of an expected drop of dividend should
lead to a proportional drop in price. Amplification of the drop occurs if 
the drop in dividend is also seen as an expected drop in future growth of dividends.
Then, equation (\ref{hfkaajkas}) applies and leads to a large amplification if 
$r'_d$ moves away from $r$. Consider for instance the $30 \%$ Procter \& Gamble crash
on 7-Mar-2000. P\&G said it expected fiscal third-quarter earnings per share 
to be down 10 to 11 percent from the year-ago
quarter, compared with its prior forecast for a rise of 7 to 9 percent.
Thus, the expected earnings dropped by $17-20 \%$. According to (\ref{hfkjks}),
the price should have dropped by a maximum of $20\%$. The extra $10\%$ can be 
rationalized by an amplification coming from a downward revision of the expected
earning growth rate $r_d$. Taking $r=5\%$ and a change of only $0.5 \%$ in the expected
$r'_d$ is enough to account for the additional $10 \%$. This illustrates quantitatively
the large susceptibility expressed by (\ref{jfjlala},\ref{jfhjamama}).

\subsection{Finite time horizon}

The formulas (\ref{hfkjks},\ref{hfkaajkas},\ref{fjjlalaZ}) assume that investors
have an infinite time horizon. In practice, humans are mortal and have a time horizon
to estimate the value of a company of the order of years to ten years or more.
It is thus interesting to quantify the effect of this finite-time effect, as 
it also allows us to get finite prices even in the phase $r \leq r'_d$, using
the firm-foundation valuation theory.
 The simplest
model is to assume that the earnings grow from year $0$ to year $Y$ at the fixed
growth rate $r_d$
and then stay constant at the value attained at year $Y$. The valuation
formula (\ref{hfkaajkas}) is then changed into
\bea
p/d  &=& 1 + {1+r'_d \over 1+r} + { (1+r'_d)^2 \over (1+r)^2} + ... +
{ (1+r'_d)^Y \over (1+r)^Y} + { (1+r'_d)^Y \over (1+r)^Y} 
\left( {1 \over 1+r} + {1 \over (1+r)^2} + ... \right)  \nonumber \\
&=& K {1+ r \over r} + {1 + r - K(1+r'_d) \over r - r'_d}~,
\eea
where
\be
K = \left({1+r'_d \over 1+r}\right)^Y~.
\ee

\begin{figure}
\begin{center}
\epsfig{file=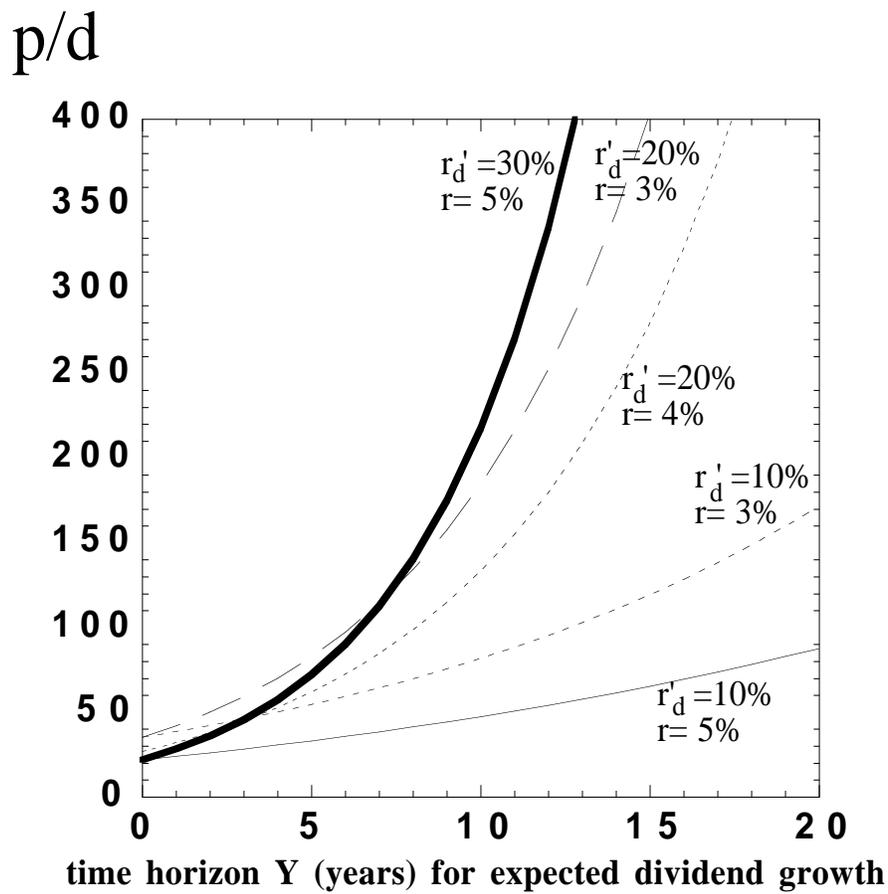,height=12cm,width=12cm}
\caption{\protect\label{P/Eyear}  Dependence of the price-over-dividend ratio $p/d$
as a function of the time horizon $Y$ over which the earnings are expected
to grow at a risk-adjusted rate $r'_d$ and then stabilize at the value at year $Y$. The five
curves show the sensitivity of $p/d$ to the discount rate $r$ and the earning
growth rate $r'_d$.
}
\end{center}
\end{figure}

Figure \ref{P/Eyear} shows that a $p/d$ of 200 can be reach for instance with 
a growth rate $r'_d=30\%$ with $r=5\%$ and a time horizon $Y \approx 9$ years
or $r'_d=20\%$ with $r=4\%$ and a time horizon $Y \approx 10$ years. 
For much more moderate growth rate $r'_d=10\%$, a similar $p/d$ requires
a small discount rate $r=3\%$ and a longer time horizon $Y > 20$ years.
This set of curves also show the large sensitivity of the price-over-dividend ratio
with respect to 
changes in $r-r'_d$ as well as time horizon $Y$, especially for the larger growth rates.
For $r'_d = 20\%$, we see the impact of an increase of discount rate from $3\%$ to $4\%$:
a major crash of more than $30\%$ in the stock price.

\subsection{Delayed expectations}

\begin{figure}
\begin{center}
\epsfig{file=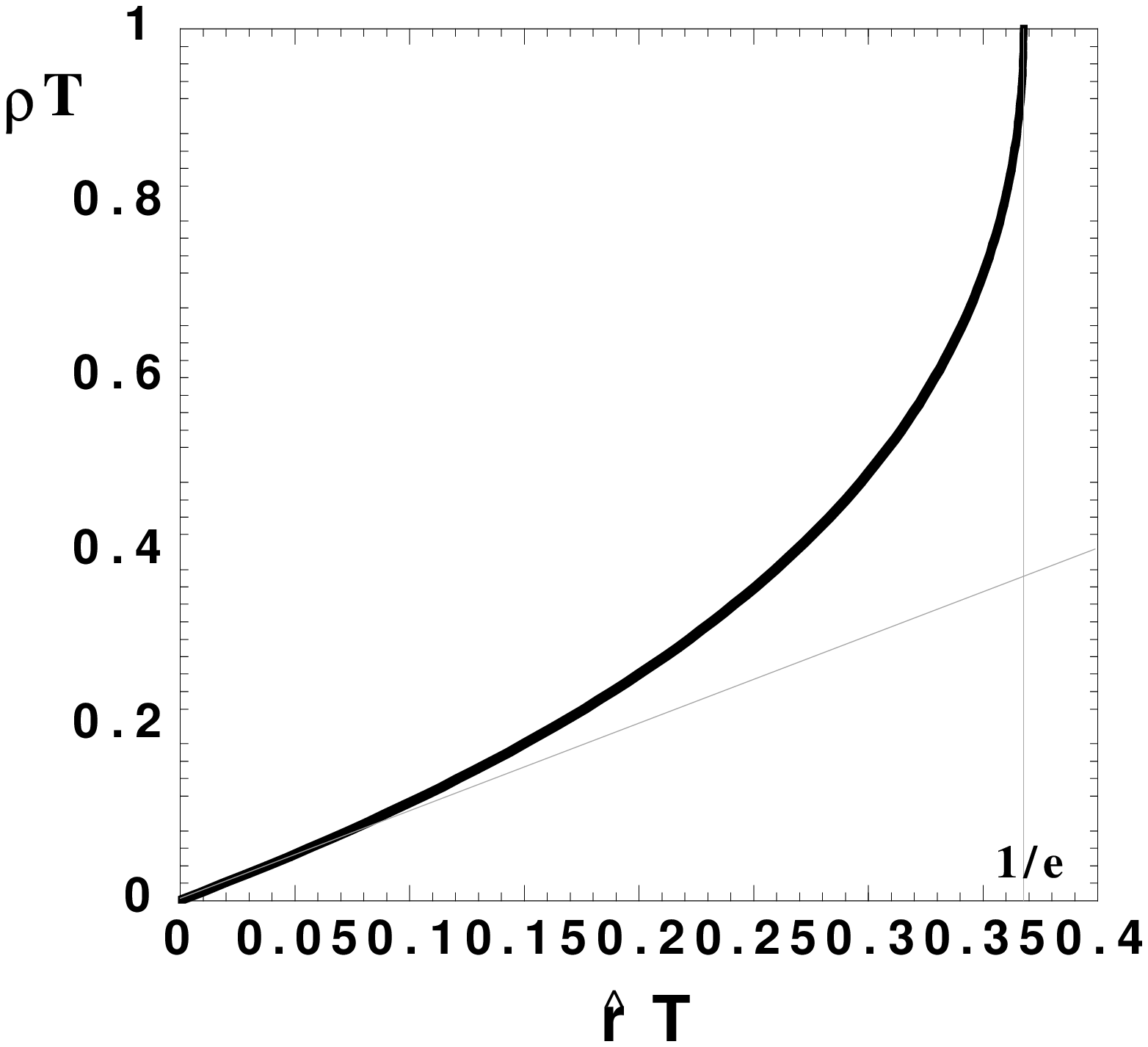,height=12cm,width=12cm}
\caption{\protect\label{delayeq}  Solution of equation (\ref{quuyquq})
showing the renormalized growth rate $\rho$ as a function of the bare
growth rate ${\hat r}$ with dimensionless units such that, by posing
$y=\rho T$ and $x={\hat r} T$, equation (\ref{quuyquq}) becomes $y=x e^y$.
Note that $\rho$ is always larger than ${\hat r}$ in the range where there is a real solution,
i.e. for ${\hat r}T \leq 1/e \approx 0.36$. 
For small ${\hat r}T$, $\rho$ is very close to ${\hat r}$ and reaches its maximum equal to $1/T$
at ${\hat r}T = 1/e$ with an infinite slope: $\rho T = 1 - C (1/e - {\hat r}T)^{-1/2}$ where
$C$ is a constant.
}
\end{center}
\end{figure}

Our theory also rationalize the real estate speculative bubbles for which
rents ($d$) had not increased significantly, but investors and buyers thought
that the price would grow without end. This spontaneous over-valuation can be make
more precise by the following future expectation equation:
\be
{dp \over dt} = {\hat r} p(t+T(t))~.   \label{fhaaaz}
\ee
Expression (\ref{fhaaaz}) expresses that the price increment has the usual 
proportionality dependence on price in the growth regime, 
however not on present but on the future price
at a time $T(t)$ in the future. If $T(t)$ is fixed, it is easy to solve this equation
and find that $p(t)$ grows exponentially but with a growth rate $\rho$ larger than ${\hat r}$ given
by the solution of the equation
\be
\rho = {\hat r} ~e^{\rho T}~,   \label{quuyquq}
\ee
obtained by looking at exponentially growing solutions in (\ref{fhaaaz}).
Since the exponential is always larger than $1$ for positive arguments ${\hat r} T$, this shows
that $\rho$ is always larger than ${\hat r}$ for positive $T$ as shown in figure 
\ref{delayeq}: the expectation of future growth 
projected a time $T$ in the future enhanced the price growth rate. It is easy to solve
iteratively this equation. There is a bifurcation point: for ${\hat r} T > 1/e$, $\rho$ develops
an imaginary part, i.e. the growth becomes oscillatory. If the optimism of investors
increases, we can capture this increase by a larger horizon time $T$ and get 
a further accelerating growth rate of the price. In this model, more confidence
in the long term growth  and thus in the long term horizon directly translates into
an accelerating growth rate as since in figure \ref{delayeq}. 
Many variations of this model (\ref{fhaaaz}) can be 
developed that we leave for a future report.

\section{Concluding remarks}

We have proposed a simple theoretical framework to account for
paradoxical behavior of the ``new economy'', based on the 
valuation of dividend growth rate. The basic concept is as follows.
If the dividend $d$ is small, the price
is large only when we take into account all the discounted future earnings
with a large growth rate $r_d \geq r$. The corresponding divergence of the price
occurs only if investors have an infinite horizon. For a finite time horizon, say
ten years, the price will be finite but will exhibit a sensitive dependence 
on the estimation of the risk-adjusted earning growth rate $r'_d$ and of the horizon up to which
this growth rate will be sustained, as seen in the previous section.
A move of the Federal Reserve to increase its basic rate $r$ will have
a strong effect in certain region of the parameter space as it will bring back
the price to lower $p/d$ ratio.

The simple valuation equation (\ref{hfkaajkas})
and its generalizations (\ref{jfjaka},\ref{jfllbala}) and (\ref{fjjlalaZ}) accounts
for a number of stylized facts observed during speculative bubbles:
\begin{itemize}
\item the sentiment that the ``run'' will last indefinitely;

\item the large increase in the price-over-dividend ratio;

\item each speculative move has had its ``growth companies'';
in 1857, the railways; in 1929, the utilities (electricity production);
in the 1960's, the office equipment companies (e.g. IBM) and
the rubber companies (car makers); today, we have the internet, software, banks and
investment companies. These companies have a fast growth rate (usually larger than 30\% per year)
and investors thus expect a large growth rate $r_d$ of their earnings.

\item Speculative phases are often stopped by successive increase of the
discount rate; this occurred in 1929 (increase fom 
3.25\% up to 6\%), in 1969, and in 1990 in Japan (increase from 
2.5\% to 6\%). 

\item The high sensitivity of valuation close to the critical point $r-r'_d=0$ 
and the spontaneous speculative valuation below it suggest that crashes and rallies
can also be interpreted as reassessments of expected risk-ajusted returns and their
growth rates.

\end{itemize}

This leads to the following avenue for future research:
new technologies, such as internet, wireless communication, wind
power, etc. should be compared to 
old technologies, such as cars, shipping, mining, etc. We expect
that stocks in the new technology class have high prices and
low earnings and thus high price-over-dividend and price-over-earning
ratios, while stocks in the old
technology class have lower prices and higher earning and then lower
price-over-dividend and price-over-earning
ratios. If one goes back in time, present ``old technology'' was new technology
and a similar pattern of high price-over-dividend and price-over-earning
ratios should be seen.
This is an interesting empirical investigation that we intend to report
on in future work.

\vskip 0.5cm
I acknowledge stimulating discussions with and suggestions from 
D. Darcet, A. Johansen, W.I. Newman, B. Roehner and D. Zajdenweber. I thank
A. Johansen for bringing my attention to the three company crashes discussed here.

\newpage

\newpage

\newpage

\newpage

\newpage

\newpage

\newpage

\begin{thebibliography}{}

\bibitem{Lahsor} J. Laherr\`ere and D. Sornette,
Stretched exponential distributions in Nature and Economy: ``Fat tails''
with characteristic scales, European Physical Journal B 2, 525-539 (1998)

\bibitem{outlier} A. Johansen and D. Sornette,
Stock market crashes are outliers, European Physical Journal B 1, 141-143 (1998)

\bibitem{White} White, E.N., ed., Stock market crashes and speculative manias, 
The international library of macroeconomic and financial history, vol. 13
(An Elgar Reference Collection, Cheltenham, UK; Brookfield, US, 1996)

\bibitem{JSL} Johansen, A., D. Sornette and O. Ledoit,
Predicting Financial Crashes using discrete scale invariance,
Journal of Risk 1 (4), 5-32 (1999); A. Johansen and D. Sornette,
Critical Crashes, Risk, Vol 12, No. 1, p.91-94 (1999).

\bibitem{Guild} Guild S.E., Stock growth and discount tables (Financial publishers, 1931).

\bibitem{Williams} Williams J.B., The theory of investment value (Harvard University Press, 1938).

\bibitem{Malkiel} Malkiel B.G., A random walk down Wall Street 
(W.W. Norton \& Company, New York, 1999)

\bibitem{Ifisher} Fisher I., The theory of interest (Kelley, 1961).

\bibitem{GrahamDodd} Graham B. and D.L. Dodd, Security analysis, 1st edition (McGraw-Hill, 1934).

\bibitem{Debreu} Debreu, G.,
Theory of value : an axiomatic analysis of economic equilibrium 
(New Haven : Yale University Press, 1987, 1959).   

\bibitem{Lamont} O. Lamont, Earnings and expected returns,
The Journal of Finance LIII, 1563-1587 (1988)

\bibitem{Gorshap} Gordon, M.J. and E. Shapiro, Capital investment analysis: the 
required rate of profit, Management Sci. 3, 102-110 (1956).

\bibitem{vonneumann} Von Neumann, J.,
"Uber ein \"okonomisches Gleichungssystem und eine Verallgemeinerung des
Brouwerschen Fixpunktsatzes (1938) 
english translation:  A model of general economic equilibrium, 
``Readings in mathematical economics'', Peter Newman editor, 
John Hopkins Press, Baltimore, 1968, p.221-229.

\bibitem{Zajbook} Zajdenweber, D., Economie des extr\^emes,  
collection Nouvelle Biblioth\`eque Scientifique (Flammarion Editor, Paris, Feb. 2000).

\bibitem{Blanchfish} Blanchard, O. and S. Fischer,
Lectures on macroeconomics (Cambridge, Mass. : MIT Press, 1989).

\bibitem{Galbraith} J.K. Galbraith, 
The great crash, 1929 (Boston: Houghton Mifflin Co., 1997).

\bibitem{kestenreview}  Sornette, D., 
Linear stochastic dynamics with nonlinear fractal properties, 
Physica A 250, 295-314 (1998).

\bibitem{Champenowne} D.G. Champenowne, {\it Economic Journal} {\bf 63}, 318-51 (1953).

\bibitem{Haan} de Haan, L., Resnick, S.I., Rootz\'en, H. \&  de Vries,
C.G. {\it Stochastic Processes and Applics.} {\bf 32}, 213-224 (1989)

\bibitem{Sorcont} Sornette, D. and R. Cont,
Convergent multiplicative processes repelled from zero: power laws and
truncated power laws, (cond-mat/9609074), J. Phys. I France 7, 431-444 (1997)

\bibitem{luxsor} Lux, T. and D. Sornette,
On Rational Bubbles and Fat Tails, submitted to the Journal 
of Money, Credit and Banking
(preprint at http://xxx.lanl.gov/abs/cond-mat/9910141)

\bibitem{calanetal} de Calan,~C., Luck,~J.-M., Nieuwenhuizen,~T.M. and
Petritis,~D. (1985) On the distribution of a random variable occurring in 1D disordered systems,
J. Phys. A 18,~501-523.

\bibitem{Bachelier} Bachelier, M.L., {\em Th\'eorie de la Sp\'eculation}
(Gauthier-Villars, Paris, 1900).

\bibitem{Samuelson} Samuelson, P.A., {\em Collected Scientific Papers} 
(M.I.T. Press, Cambridge, MA, 1972).

\bibitem{technianaly} Andersen, J. V., S. Gluzman and D. Sornette,
Fundamental Framework for Technical Analysis, European Physical Journal B 14, 579-601 (2000).

\bibitem{Staubook} Moss de Oliveira, S., P.M.C de Oliveira and D. Stauffer,
{\em Evolution, Money, War and Computers}, Teubner, Stuttgart-Leipzig, 1999.

\bibitem{Goldenfeld} Goldenfeld,~N. (1992) {\it Lectures on Phase
Transitions and the Renormalization Group} (Addison-Wesley,~Advanced
Book Program,~ Reading,~Mass.).

\bibitem{SSBrefs} J. Sivardiere,
Spontaneous symmetry breaking in a Cavendish experiment,
American Journal of Physics 65, 567-568 (1997);
P.K. Aravind, A simple geometrical model of spontaneous symmetry breaking,
American Journal of Physics 55, 437-439 (1987);
J.R. Drugowich de Felicio and O. Hipolito, 
Spontaneous symmetry breaking in a simple mechanical model,
American Journal of Physics 53, 690-693 (1985)
J. Sivardiere, A simple mechanical model exhibiting a spontaneous symmetry breaking,
American Journal of Physics 51, 1016-1018 (1983);
M. Consoli and P.M. Stevenson, 
Physical mechanisms generating spontaneous symmetry breaking and a
hierarchy of scales, Int. J. Mod. Phys. A 15, 133-157 (2000).
       
\bibitem{Weinberg} S. Weinberg, 
The quantum theory of fields (Cambridge; New York:
Cambridge University Press, 1995-1996).

\end{thebibliography}
\end{document}